\documentclass[useAMS,usenatbib]{mn2e}
\usepackage{amsmath,amssymb}
\usepackage[papersize={8.5in,11in}]{geometry}

\newcommand{\be}{\begin{equation}}
\newcommand{\ee}{\end{equation}}

\begin{document}

\title{NGC 1052-DF2 And Modified Gravity (MOG) Without Dark Matter}

\author[J. W. Moffat and V. T. Toth]{J. W. Moffat$^1$ and V. T. Toth$^1$\\
$^1$Perimeter Institute for Theoretical Physics, Waterloo, Ontario N2L 2Y5, Canada}

\maketitle

\begin{abstract}
We model the velocity dispersion of the ultra-diffuse galaxy NGC 1052-DF2 using Newtonian gravity and modified gravity (MOG). The velocity dispersion predicted by MOG is higher than the Newtonian gravity prediction, but it is fully consistent with the observed velocity dispersion that is obtained from the motion of 10 globular clusters (GCs).
\end{abstract}


\maketitle

\section{Introduction}

Observations of groups of clusters-like objects in the diffuse galaxy NGC 1052-DF2 using the Dragonfly Telephoto Array, the Sloan Digital Sky Survey data, the Gemini Observatory, the Hubble Space Telescope and the 10~m W. M. Keck observatory have revealed that the velocity dispersion of NGC 1052-DF2 is consistent with its measured stellar mass without dark matter~\citep{Dokkum1,Dokkum2}.  The stellar mass of this galaxy, estimated at $M_\star\sim 2\times 10^8~M_\odot$, is consistent with its velocity dispersion, which is measured at $<10.5$~km/s with $90\%$ confidence. From this it is inferred that NGC 1052-DF2 contains little or no dark matter. This is the first galaxy that has been observed to lack dark matter.

We demonstrate that this lack of dark matter is consistent with the modified gravity theory Scalar-Tensor-Vector-Gravity (STVG), also known as MOG~\citep{Moffat1}, using parameter choices that were established in earlier studies of, e.g., spiral galaxies. The theory fits globular clusters~\citep{MoffatToth1}, a large selection of galaxies without dark matter~\citep{Moffat1,MoffatRahvar1,MoffatToth2,MoffatToth3}, as well as galaxy clusters~\citep{MoffatRahvar2,MoffatBrownstein2,MoffatIsrael}. It can also explain cosmology data~\citep{MoffatToth3}, and it is consistent with the recent gravitational wave and multispectral observation of GW170817 \citep{MoffatGreenToth}. The detection of a galaxy with little or no indication of the presence of dark matter that nonetheless behaves in a manner consistent with the predictions of MOG strengthens the conclusion that dark matter is not detectable in the present universe and that the gravitational theory of Newton and Einstein requires further modification.

\section{MOG Acceleration}

The MOG modified acceleration law for a point source in weak gravitational fields and low velocities, derived from the MOG Lagrangian and field equations and the test particle equation of motion is given by \citep{Moffat1}:
\begin{equation}
\label{acceleration}
a(r)=-\frac{G_NM}{r^2}[1+\alpha-\alpha\exp(-\mu r)(1+\mu r)],
\end{equation}

The parameter $\alpha$ can be determined from~\citep{MoffatToth1}:
\begin{equation}
\label{alpha}
\alpha_{\infty}=(G_{\infty}-G_N)/G_N,
\end{equation}
where $G_\infty$ is the asymptotic limit of $G$ for very large mass concentrations. This yields
\be
\alpha=\alpha_{\infty}\frac{M}{(\sqrt{M}+E)^2}.
\ee
Moreover, we have for the parameter $\mu$:
\begin{equation}
\label{mu}
\mu=\frac{D}{\sqrt{M}}.
\end{equation}
Here, $D$ and $E$ are given by \citep{MoffatToth5}:
\begin{align}
D&=6.25\times 10^3\,{M_{\odot}}^{1/2}\,{\rm kpc}^{-1},\\
E&=2.5\times 10^4\,{M_{\odot}}^{1/2}.
\end{align}

We choose $\alpha_\infty=10$. This is consistent with the best current estimate for the baryonic mass of the Milky Way, $M_b^{\rm MW}\sim 1.7\times 10^{11}~M_\odot$, which yields $\alpha^{\rm MW}=8.89$, the value used in our earlier study of Galaxy rotation curves~\citep{MoffatRahvar1,MoffatToth2}.

From (\ref{alpha}) and (\ref{mu}) for the mass of NGC 1052-DF2 $\sim 2\times 10^8\,{M}_{\odot}$ and $\alpha_{\infty}=10$ we obtain the values
\begin{equation}
\alpha=1.30,\quad \mu=0.443\,{\rm kpc}^{-1}.
\end{equation}

\section{Velocity dispersion for spherically symmetric systems}

We investigate systems that are characterized by a two-dimensional S\'ersic profile, parameterized by the S\'ersic index $n$, axis ratio $b/a$ and effective radius along the major axis $R_e$.

To keep the problem tractable without a serious loss of precision, we scale the system to a spherical S\'ersic profile with index $n$ and effective radius ${R}_e\to\frac{1}{2}(1+b/a)R_e$. The surface brightness of such a system is given by
\be
I(R)=I_0\exp\left(-\beta[R/R_e]^{1/n}\right),
\label{eq:I(R)}
\ee
where $I_0$ is the surface brightness per unit surface area, whereas\footnote{This truncated approximation is accurate to $10^{-3}$ for $n=0.6$, used later in this paper.} $\beta\sim 2n-1/3+4/405m+46/25515m^2+{\cal O}(m^{-3})$ \citep{CiottiBertin}. The integrated luminosity that corresponds to this profile is given by simple integration:
\be
2\pi\int_0^\infty RI(R)~dR=\pi\Gamma(1+2n) I_0 \beta^{-2n} R_e^2.
\label{eq:Sigma}
\ee
If the total luminosity of the object is known, this integral can be used to obtain the value of $I_0$.

The surface brightness of a {\em transparent} spherically symmetric galaxy with {\em luminosity density} $j(r)$ is given by \citep{BT2008}:
\be
I(R)=2\int_R^\infty dr\frac{rj(r)}{\sqrt{r^2-R^2}}.
\ee
Thus \citep{BT2008},
\be
j(r)=-\frac{1}{\pi}\int_r^\infty\frac{dR}{\sqrt{R^2-r^2}}\frac{dI}{dR}.
\label{eq:j(r)}
\ee
If we assume that the mass-to-light ratio of the galaxy is constant (no dependence on $R$),
\be
\rho(r)=\Upsilon_0 j(r),
\label{eq:rhodef}
\ee
we can write down formally identical expressions for the mass and mass density of the galaxy. For Newtonian gravity, radial acceleration in a spherically symmetric mass distribution is given by
\be
a(r)=-\frac{G_N}{r^2}\int_0^r4\pi\rho(r')r'^2~dr',
\ee
as any mass density outside $r$ need not be included, since an inverse-$r^2$ gravity theory satisfies the shell theorem.

In MOG, the acceleration in a spherically symmetric mass distribution can be calculated as
\begin{align}
a(r)=&
-\int_0^{r}dr'\dfrac{2\pi Gr'}{\mu r^2}\rho(r')\bigg\{2(1+\alpha)\mu r'\\
&{}+\alpha(1+\mu r)\left[e^{-\mu(r+r')}-e^{-\mu(r-r')}\right]\bigg\}\nonumber\\
-&\int_{r}^\infty dr'\dfrac{2\pi Gr'}{\mu r^2}\rho(r')\alpha\nonumber\\
&{}\times\bigg\{\big[1+\mu r\big]e^{-\mu(r'+r)}-\big[1-\mu r\big]e^{-\mu(r'-r)}\bigg\}.\nonumber
\end{align}

The velocity dispersion for a spherical system with no velocity anisotropy is given by the Jeans equation in the form
\be
\frac{\partial(\rho\sigma^2)}{\partial r}-\rho a=0,
\ee

Using $\lim\limits_{r\rightarrow\infty}\sigma^2(r)=0$, we get the solution by integration:
\be
\sigma^2(r)=\frac{1}{\rho(r)}\int_r^{\infty}\rho(r') a(r')~dr'.
\label{eq:s2}
\ee
To obtain the line-of-sight root-square average of $\sigma^2(r)$ for a tracer population of $N$ objects characterized by a number density $\rho_N$, we calculate
\be
\sigma^2_{\|}=\frac{4\pi}{N}\int\rho_N(r')\sigma^2(r')r'^2~dr'.
\label{eq:spar}
\ee

\section{The case of NGC 1052-DF2}

The ultra-diffuse galaxy NGC 1052-DF2 is modeled using a two-dimensional S\'ersic profile, characterized by $n=0.6$, $b/a=0.85$, $R_e=2.2$~kpc \citep{Dokkum1}. We approximate this profile using a spherical S\'ersic profile with $R_e=2.0$~kpc.

Assuming a constant mass-to-light ratio, we describe the mass profile of NGC 1052-DF2 using (\ref{eq:I(R)}) together with (\ref{eq:rhodef}).
The corresponding total mass, using (\ref{eq:Sigma}), is
\be
M=\pi\Gamma(2.2)\Sigma_0\beta^{-2n}R_e^2.
\ee
The stellar mass estimate of NGC 1052-DF2 is $M_\star=2\times 10^8~M_\odot$, which we take to represent its baryonic mass $M$, from which we calculate its characteristic surface mass density as
\be
\Sigma_0=\frac{M}{\pi\Gamma(2.2)\beta^{-2n}R_e^2}=1.25\times 10^7~M_\odot/{\rm kpc}^2.
\ee
The radial mass density given by (\ref{eq:j(r)}) in conjunction with (\ref{eq:rhodef})
can, in fact, be calculated in closed form using Meijer's G-function. This closed form unfortunately is not very useful in further calculations. The mass density can, however, be closely approximated as
\be
\rho(r)\sim\frac{40\Sigma_0}{63R_e}\exp\left(-\left[\frac{11r}{10R_e}\right]^{4/3}\right),
\ee
with the numerically evaluated error not exceeding 10\% for $R\lesssim 7$~kpc.

Finally, we represent the tracer population of 10 GCs used in \citep{Dokkum1,Dokkum2} by way of a simple exponential fit, $\rho_N\propto\exp(r^2/r_0^2)$, with a best fit obtained for $r_0=4.46$~kpc. Numerically evaluating (\ref{eq:s2}) and (\ref{eq:spar}) using Newtonian gravity yields
\be
\sigma^2_{\|_{\rm Newton}}=(2.7~{\rm km}/{\rm s})^2,
\ee
whereas for MOG, we obtain
\be
\sigma^2_{\|_{\rm MOG}}=(3.9~{\rm km}/{\rm s})^2.
\ee
Both these values agree well with the reported intrinsic velocity dispersion, $\sigma_{\rm intr}=3.2^{+5.5}_{-3.2}~{\rm km}/{\rm s}$~\citep{Dokkum1}. The results are moderately sensitive to the fitted parameter $r_0$ characterizing the somewhat arbitrarily chosen exponential density model. Varying $r_0$ by $\pm$50\% yields $(2.0~{\rm km}/{\rm s})^2\le\sigma^2_{\|_{\rm Newton}}\le(3.8~{\rm km}/{\rm s})^2$ and $(2.9~{\rm km}/{\rm s})^2\le\sigma^2_{\|_{\rm MOG}}\le(5.3~{\rm km}/{\rm s})^2$.

\section{Conclusions}

Straightforward application of the MOG/STVG acceleration law, first published in its present form in 2006~\citep{Moffat1}, yields a velocity dispersion for the recently studied ultra-diffuse galaxy NGC 1052-DF2~\citep{Dokkum1,Dokkum2} that is in good agreement with observation. Thus, contrary to claims that suggested otherwise, the case of NGC 1052-DF2 does not contradict modified gravity as an alternative to dark matter; and specifically, it is not in any tension with the predictions of our MOG/STVG theory. To the contrary, in the absence of any viable mechanism that would separate {\em stellar} mass from collisionless dark matter to reduce the dark matter ratio to acceptable levels, one has to conclude that NGC 1052-DF2 is a challenge to the prevailing dark matter paradigm.

This raises the question: What about other ultra-diffuse galaxies? Galaxies with masses similar to that of NGC 1052-DF2 but velocity dispersions in excess of 30~km/s, or galaxies that have velocity dispersions similar to that of NGC 1052-DF2, but with stellar masses that are more than an order of magnitude smaller? While we recognize this as a challenge to modified gravity, we also note that the case of such objects is far from being settled: that dark matter halos often fare no better than modified gravity ~\citep{Kroupa2012}, indicating that many of such systems may not be fully virialized, and, as such, may be poor candidates for testing gravitational dynamics, possibly due to their history, for instance, encounters with massive host galaxies of which they are satellites. If this is indeed the case, NGC 1052-DF2 may represent one of the exceptions: an ultra-diffuse galaxy that is free of disruptions, and demonstrates almost Newtonian behavior with no indication of a massive dark component. The accuracy of velocity dispersion measurements has also been recently brought into question~\citep{Martin,Navarro2018}; the selection of tracers and other sources of bias and uncertainty can result in estimates of large dark matter halos or no dark matter halos at all for the same ultra-diffuse galaxy.

Notwithstanding the above, we demonstrated that MOG/STVG reproduces the observed velocity dispersion of NGC 1052-DF2 with ease, requiring no additional assumptions and no changes in the theory's parameterization when compared with previous studies.

\section*{Acknowledgments}

This research was supported in part by Perimeter Institute for Theoretical Physics. Research at Perimeter Institute is supported by the Government of Canada through the Department of Innovation, Science and Economic Development Canada and by the Province of Ontario through the Ministry of Research, Innovation and Science. We thank the anonymous referee for valuable contributions.

\end{document}